# Modeling Uncertainties in Aggregated Thermostatically Controlled Loads Using a State Queueing Model

N. Lu *Member, IEEE*, D. P. Chassin, *Member, IEEE*, and S. E. Widergren, *Sr. Member, IEEE*

***Abstract* — To study the impacts of price responsive demand on the electric power system requires better load models. This paper discusses the modeling of uncertainties in aggregated thermostatically controlled loads using a state queueing (SQ) model. The cycling times of thermostatically controlled appliances (TCAs) vary with the TCA types and sizes, as well as the ambient temperatures. The random consumption of consumers, which shortens or prolongs a specific TCA cycling period, introduces another degree of uncertainty. By modifying the state transition matrix, these random factors can be taken into account in a discrete SQ model. The impacts of considering load diversity in the SQ model while simulating TCA setpoint response are also studied.**

## I. Introduction

IN a competitive electricity market, increased demand elasticity can moderate suppliers' ability to exercise market power and alleviate excessive price volatility. Customer responses to the market prices can be classified into four different categories: curtailment [1], substitution [2], storage [3], and shifting. Load shifting, as the name implies, shifts electric usage to pre- or post-peak periods to reduce consumption during the anticipated peak-price periods. An important feature of the load shifting program is that it targets the cyclic loads such as thermostatically controlled appliances (TCAs). TCAs include residential heating ventilation and air conditioning (HVAC) systems, electric water heaters, and refrigerators. Varying the setting of a TCA thermostat can shift the TCA power consumption from tens of minutes to a couple of hours, depending on the appliances. If the setpoint is controlled in response to the market prices, the shifted TCA's power consumption can contribute to load reduction during the peak-price periods.

To evaluate the economic benefits of different load shifting strategies, it is essential to model the change in TCA power consumption as a function of setpoint changes in response to market prices. Load models developed for TCAs based on statistical analysis [4-7] of historical data can not model the power output of a TCA when the setpoint is controlled in response to market prices. A state queueing (SQ) model [8] was proposed by Lu and Chassin to simulate aggregated TCA loads after their setpoint changes. This paper is a follow-up paper to the initial work done in [8], where TCAs are assumed to have the same size and are set at the same setpoint under similar ambient temperatures, leading to a simplified model that only accounts for standby loss. However, in practice, TCA units have different sizes and the ambient temperatures vary in different regions. Furthermore, random customer usages will have significant impact on the power consumptions of TCAs. In this paper, using a water heater model as an example, methodologies are developed to modify the transition matrixes to account for the uncertainties in thermal parameters, the ambient temperatures, and the random customer behaviors, which are the key for the SQ model to simulate aggregated TCA loads accurately. The impacts of considering the randomness in the model on the TCA setpoint response are also studied.

The paper is organized as follows. Section II introduces the SQ model. Section III describes uncertainties in thermal model parameters and presents the modified thermal model. Customer random behaviors are discussed and modeled in Section IV. The impacts of combining the modeling of uncertainties on setpoint change response are discussed in Section V. Section VI provides conclusion.

## II. TCA Thermal Models

### A. *Thermal Characteristics of TCAs*

Thermostatically controlled appliances include residential HVAC systems, electric water heaters, and refrigerators. Fig. 1 shows the thermal behavior (temperature of stored water) of a water heater unit over time. The rising curves indicate the water heater is "on", and the falling curves represent the standby (or cooling down) periods, when the heater is "off". As the water heater unit cycles, the water temperature in the tank rises and falls accordingly [9].

The operations of other TCAs are similar. The upper and lower limits represent the dead band of the thermostat around the thermostat setpoint, and changing the setpoint allows one to regulate the power consumption of the TCAs. Because the asymptotic equilibrium temperatures are generally far beyond these limits for appropriately sized equipment, the exponential rising curve and falling curve are nearly linear between the upper limit and the lower limit, as shown in Fig. 1. To

This work is supported by the Pacific Northwest National Laboratory operated for the U.S. Department of Energy by Battelle Memorial Institute under contract DE-AC06-76RL01830.

N. Lu, D. P. Chassin, and S. E. Widergren are with the Energy Science and Technology Division, Pacific Northwest National Laboratory, P.O. Box 999, MSIN: K5-20, Richland, WA - 99352, USA (e-mail: ning.lu@pnl.gov, david.chassin@pnl.gov, steve.widergren@pnl.gov)

simplify the analysis, a linear approximation of TCA thermal characteristics (Fig. 2) is used in our model. $T_+$ and $T_-$ are the upper and lower temperature limit for a given setpoint $T$. The unit has a cycling time of $\tau$, with an "on" period of $\tau_{on}$ and an "off" period of $\tau_{off}$.

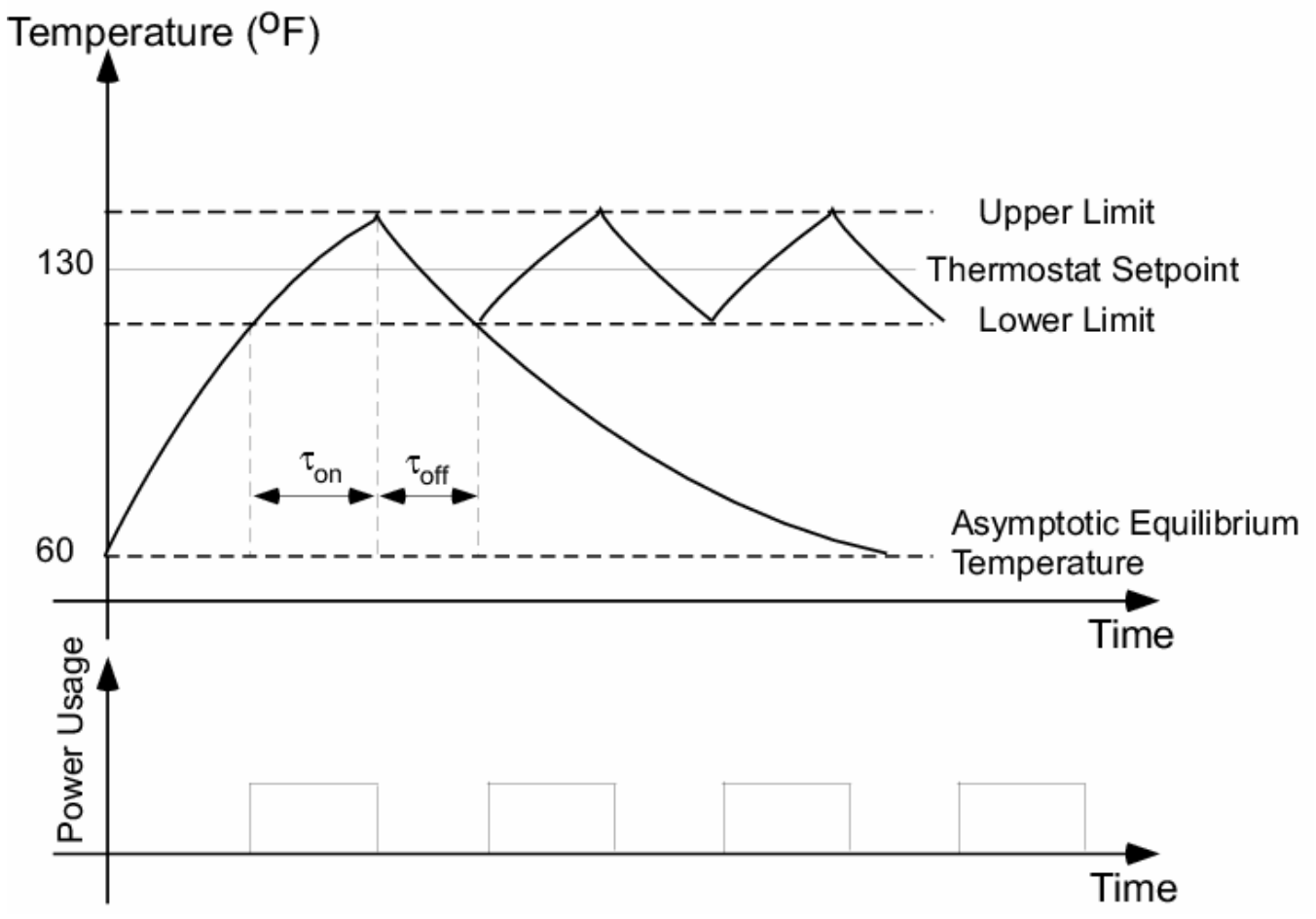

Fig. 1: Differential models of controlled thermal behavior [9]

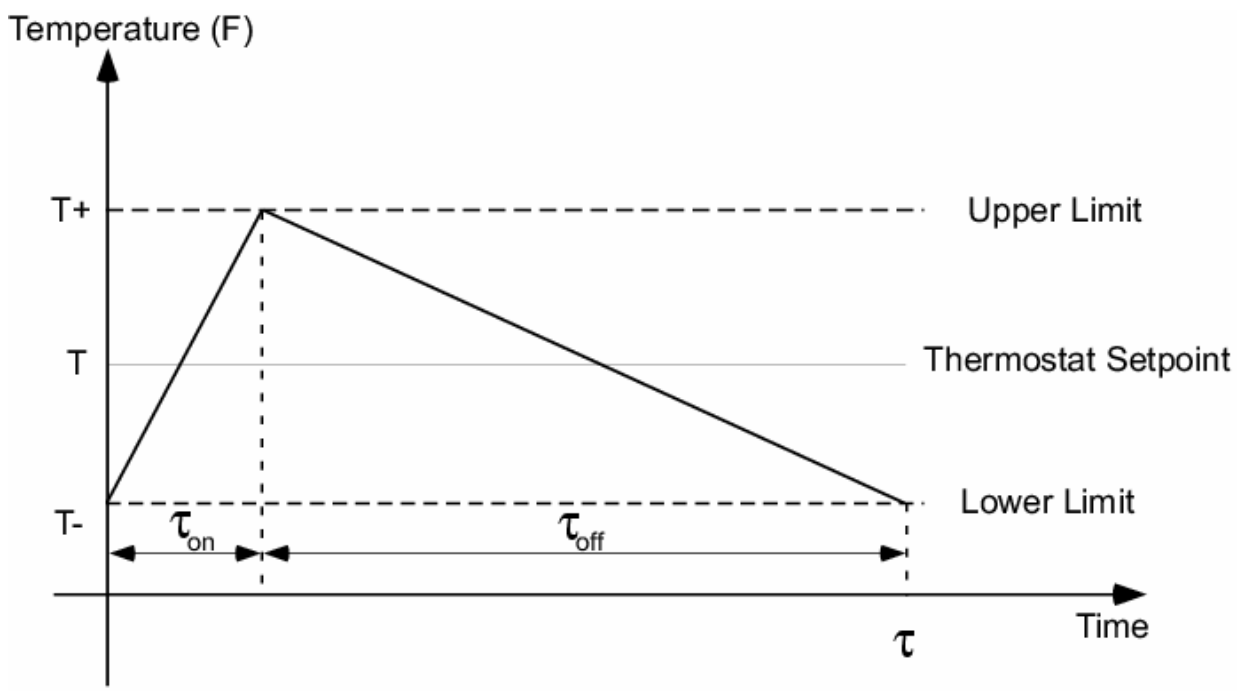

Fig. 2: The simplified thermal characteristic curve of a TCA

B. *A state queueing model* [8]

A SQ model describes a water heater behavior by tracking the state of the water heater over time. Consider a system containing $N$ water heater units with the initial thermal states shown in Fig. 3. As an illustration, consider a group of water heater units, whose "on" times are 5 minutes and "off" times are 15 minutes at a given ambient temperature and a given setpoint. Note that in practice, water heaters usually have cycling times up to 10 hours (when there is no water draw). $T_+$ and $T_-$ are the upper and lower temperature limits for a given setpoint $T$. A state is then defined by both the water temperature in the tank and the on/off status of a unit. We divide the time cycle into 20 states of equal duration such that there are 5 distinct "on" states (shown by shaded boxes in Fig. 3) and 15 distinct "off" states in a temperature range of [$T_-$, $T_+$].

Initially, we assume a uniformly diversified load, and the units are distributed uniformly among all 20 states. If the whole time period is divided into 20 time steps, then at time step 1, we will have nearly equal numbers of units in each of the 20 states. At the end of each time step, units will move one state ahead. For example, at time step 2, units in State 1 will move to State 2, units in State 2 to State 3, and units in State 20 to State 1, as shown in Table I.

Assume the average power consumption of each unit is $P_{ave}$. The total load at each time step is simply the aggregated power output of all the "on" units in States 1 through 5. When the units are uniformly distributed among the states, the expected load $P_L$ can be calculated as:

$$E(P_L) = E(N_{on})P_{ave} = \frac{n_{on}}{n} NP_{ave} = \frac{5}{20} NP_{ave}$$

where $n$ is the total number of states, $n_{on}$ is the number of "on" states, $N$ is the number of units, and $N_{on}$ is the number of "on" units.

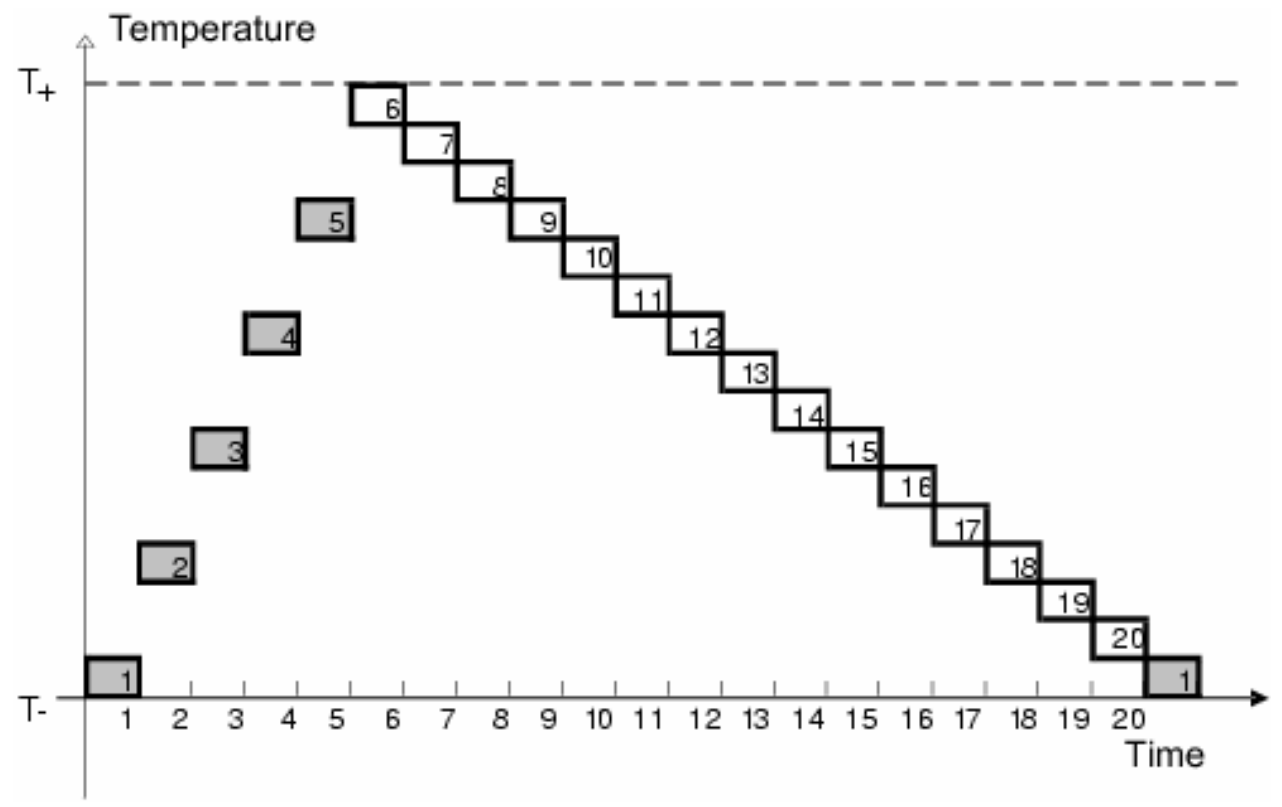

Fig. 3: The SQ model of aggregated water heater load

Table I shows the state evolution along the timeline. Rows represent time steps and columns represent the unit states. The shaded states are the "on" states. One can calculate the aggregated output by summing up the power consumption of all the units in the "on" states at each time step.

For example, if there is a 1-kW unit in each of the 20 states, the aggregated consumption at any time is 5 kW because at any time, there are 5 "on" states in the system.

TABLE I
THE STATE DISTRIBUTION OF THE WATER HEATER UNITS

| Time \ States | 1 | 2 | 3 | 4 | 5 | 6 | 7 | 8 | 9 | 10 | 11 | 12 | 13 | 14 | 15 | 16 | 17 | 18 | 19 | 20 |
|---|---|---|---|---|---|---|---|---|---|---|---|---|---|---|---|---|---|---|---|---|
| 1 | 2 | 3 | 4 | 5 | 6 | 7 | 8 | 9 | 10 | 11 | 12 | 13 | 14 | 15 | 16 | 17 | 18 | 19 | 20 | 1 |
| 2 | 3 | 4 | 5 | 6 | 7 | 8 | 9 | 10 | 11 | 12 | 13 | 14 | 15 | 16 | 17 | 18 | 19 | 20 | 1 | 2 |
| 3 | 4 | 5 | 6 | 7 | 8 | 9 | 10 | 11 | 12 | 13 | 14 | 15 | 16 | 17 | 18 | 19 | 20 | 1 | 2 | 3 |
| 4 | 5 | 6 | 7 | 8 | 9 | 10 | 11 | 12 | 13 | 14 | 15 | 16 | 17 | 18 | 19 | 20 | 1 | 2 | 3 | 4 |
| 5 | 6 | 7 | 8 | 9 | 10 | 11 | 12 | 13 | 14 | 15 | 16 | 17 | 18 | 19 | 20 | 1 | 2 | 3 | 4 | 5 |
| 6 | 7 | 8 | 9 | 10 | 11 | 12 | 13 | 14 | 15 | 16 | 17 | 18 | 19 | 20 | 1 | 2 | 3 | 4 | 5 | 6 |
| 7 | 8 | 9 | 10 | 11 | 12 | 13 | 14 | 15 | 16 | 17 | 18 | 19 | 20 | 1 | 2 | 3 | 4 | 5 | 6 | 7 |
| 8 | 9 | 10 | 11 | 12 | 13 | 14 | 15 | 16 | 17 | 18 | 19 | 20 | 1 | 2 | 3 | 4 | 5 | 6 | 7 | 8 |
| 9 | 10 | 11 | 12 | 13 | 14 | 15 | 16 | 17 | 18 | 19 | 20 | 1 | 2 | 3 | 4 | 5 | 6 | 7 | 8 | 9 |
| 10 | 11 | 12 | 13 | 14 | 15 | 16 | 17 | 18 | 19 | 20 | 1 | 2 | 3 | 4 | 5 | 6 | 7 | 8 | 9 | 10 |
| 11 | 12 | 13 | 14 | 15 | 16 | 17 | 18 | 19 | 20 | 1 | 2 | 3 | 4 | 5 | 6 | 7 | 8 | 9 | 10 | 11 |
| 12 | 13 | 14 | 15 | 16 | 17 | 18 | 19 | 20 | 1 | 2 | 3 | 4 | 5 | 6 | 7 | 8 | 9 | 10 | 11 | 12 |
| 13 | 14 | 15 | 16 | 17 | 18 | 19 | 20 | 1 | 2 | 3 | 4 | 5 | 6 | 7 | 8 | 9 | 10 | 11 | 12 | 13 |
| 14 | 15 | 16 | 17 | 18 | 19 | 20 | 1 | 2 | 3 | 4 | 5 | 6 | 7 | 8 | 9 | 10 | 11 | 12 | 13 | 14 |
| 15 | 16 | 17 | 18 | 19 | 20 | 1 | 2 | 3 | 4 | 5 | 6 | 7 | 8 | 9 | 10 | 11 | 12 | 13 | 14 | 15 |
| 16 | 17 | 18 | 19 | 20 | 1 | 2 | 3 | 4 | 5 | 6 | 7 | 8 | 9 | 10 | 11 | 12 | 13 | 14 | 15 | 16 |
| 17 | 18 | 19 | 20 | 1 | 2 | 3 | 4 | 5 | 6 | 7 | 8 | 9 | 10 | 11 | 12 | 13 | 14 | 15 | 16 | 17 |
| 18 | 19 | 20 | 1 | 2 | 3 | 4 | 5 | 6 | 7 | 8 | 9 | 10 | 11 | 12 | 13 | 14 | 15 | 16 | 17 | 18 |
| 19 | 20 | 1 | 2 | 3 | 4 | 5 | 6 | 7 | 8 | 9 | 10 | 11 | 12 | 13 | 14 | 15 | 16 | 17 | 18 | 19 |
| 20 | 1 | 2 | 3 | 4 | 5 | 6 | 7 | 8 | 9 | 10 | 11 | 12 | 13 | 14 | 15 | 16 | 17 | 18 | 19 | 20 |

time

## III. Uncertainties in the Thermal Models Of TCAs

### A. *Uncertainties in thermal model parameters*

The thermal model we used is based on an equivalent thermal parameter (ETP) approach [11, 12]. An example ETP representation of a water heater is shown in Fig. 4, where *UA* is the standby heat loss coefficient, *C* is the tank capacity, $T_{out}$ is the ambient temperature, and *Q* is the heat rate of the water heater. *Q* is proportional to the heater rated power.

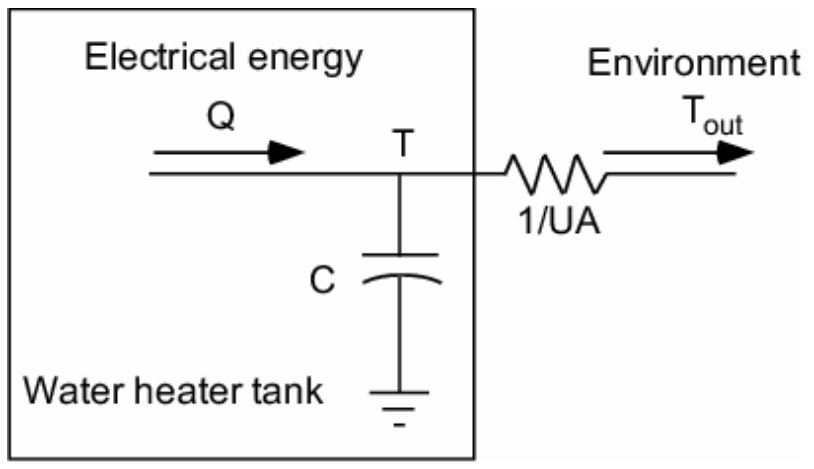

Fig. 4: An ETP heat transfer model for a water heater

Based on the ETP model, starting from an initial temperature $T_0$ and ignoring the hot water consumption, the water temperature *T* can be calculated by

$$T = (T_0 - \frac{K_2}{K_1})e^{-K_1 t} + \frac{K_2}{K_1} \quad (1)$$

where $K_1$ and $K_2$ are calculated by

$$K_1 = \frac{UA}{C} \qquad K_2 = Q + T_{out}\frac{UA}{C}$$

To calculate $\tau_{on}$, assuming $T_0 = T_-$ and $T = T_+$, we have

$$\tau_{on} = -\ln\left(\frac{T_+ - \frac{K_2}{K_1}}{T_- - \frac{K_2}{K_1}}\right)/K_1 = -\ln\left(\frac{T_+ - \frac{Q}{K_1} - T_{out}}{T_- - \frac{Q}{K_1} - T_{out}}\right)/K_1 \quad (2)$$

When the heater is turned off, *Q* is zero. The unit then coasts from $T_+$ to $T_-$. To calculate $\tau_{off}$, we have

$$Q = 0 \quad \Rightarrow K_2 = T_{out}\frac{UA}{C}$$

$$\tau_{off} = -\ln\left(\frac{T_- - \frac{K_2}{K_1}}{T_+ - \frac{K_2}{K_1}}\right)/K_1 = -\ln\left(\frac{T_- - T_{out}}{T_+ - T_{out}}\right)/K_1 \quad (3)$$

Based on (2) and (3), a sensitivity analysis can be performed to study the impact of parameter uncertainties on TCA cycling times. The analysis can be readily extended to the HVAC and the refrigerator models, which are similar to the water heater model.

#### 1) *The uncertainties in UA*

The value of *UA* varies even for the same type and size water heaters. To account for the variations, *UA* can be considered to follow a certain probability density function (PDF) in [$UA_-$, $UA_+$], such as a uniform or a normal distribution.

The heat rate of a water heater *Q* usually is much greater than the heat losses by design. Therefore, the value of *UA* will have a minor influence on the rising curve ($\tau_{on}$), where the heater is on. When the heater is "off", *Q* is zero. Then, the heat loss determined by the *UA* of the water heater will dominate the length of the falling curve ($\tau_{off}$).

As an illustration, consider 40 gallon water heaters rated at 4.5 kW with their *UA* varying from 2.4 to 3.6 Btu/hr°F. If the ambient temperature is 60°F, the calculated standby cycling periods are shown in Fig. 5. As predicted, the $\tau_{on}$ values are almost the same for all units, but the $\tau_{off}$ values scatter from 12 hr to 18 hr.

At the feeder end, where the individual load aggregates, the differences in water heater cycling times caused by different *UA* of each unit result in an uncertainty in the aggregated thermal characteristic curve used to set up a SQ model. However, because *UA* is a deterministic parameter for each water heater unit, its impact is time invariant.

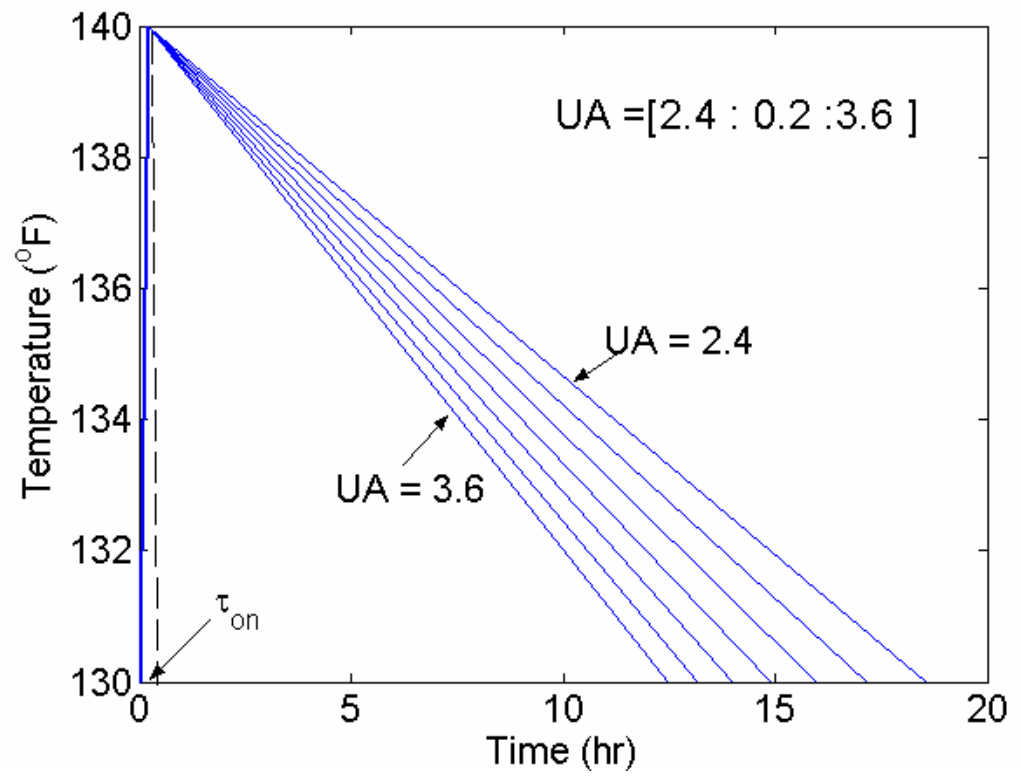

Fig. 5: The thermal characteristic curves of water heaters with different *UA*

#### 2) *Uncertainties caused by different ambient temperature*

Fig. 6 shows an example of 40 gallon water heaters rated at 4.5 kW with *UA* at 3 Btu/hr°F. The ambient temperatures vary diurnally and seasonally from 30°F to 60°F. Note that this variation is a temporal variation, and its influence is a function of time. Again the $\tau_{on}$ values change slightly while the $\tau_{off}$ values scatter around a range of a few hours.

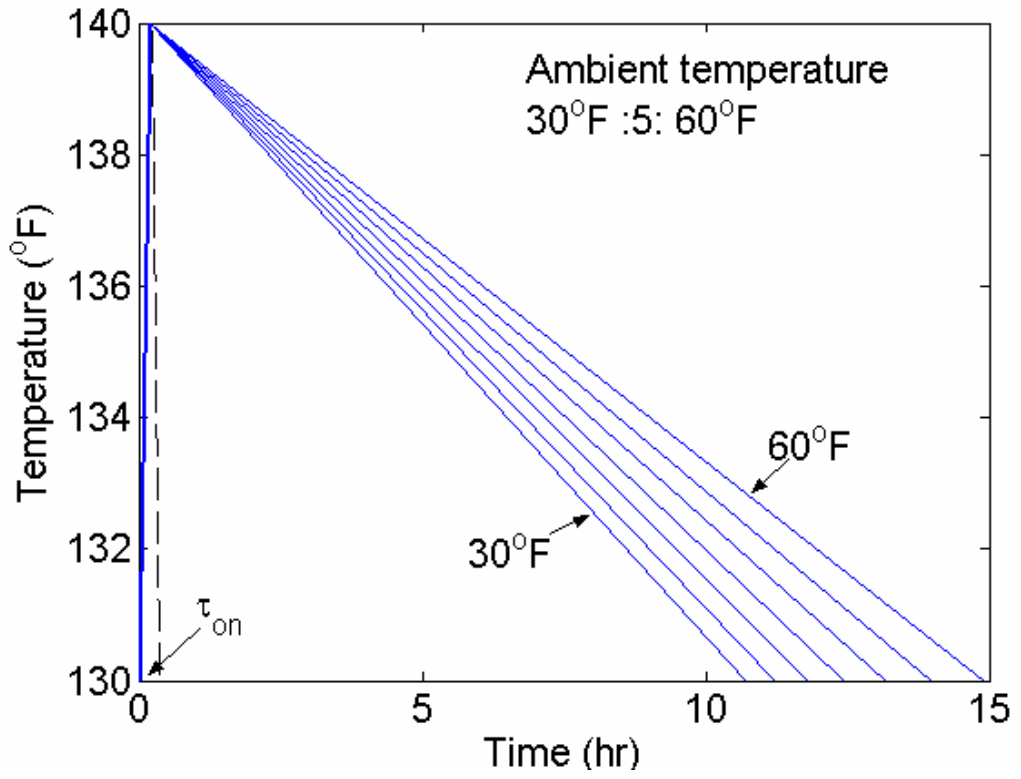

Fig. 6: The thermal characteristic curves of water heaters with different ambient temperature

This is because for water heaters, the dead band is small compared with the value of $T$-$T_{out}$. Therefore, based on (2), $\tau_{on}$ will not change significantly with different $T_{out}$. Because the ambient temperatures vary with time, the uncertainties are time dependent.

### *B. Modified thermal model of TCAs*

Taking into account the above uncertainties, the thermal characteristic curves for a group of TCAs are shown in Fig. 7. The state that a unit resides in follows some probability distribution function determined by the uncertainties brought by the different UAs and ambient temperatures. When the ambient temperatures are high, there are more units having longer cycling periods and vice versa. The probability distribution will then be shifted accordingly, as shown in Fig. 7. The uncertainties bring additional off-diagonal state transition probabilities in the probability transition matrix $P$, as shown in Fig. 8. A SQ model considering the parameter uncertainties is shown in Fig. 9, where $N$ is the total number of states, and $N_{on}$ is the number of "on" states.

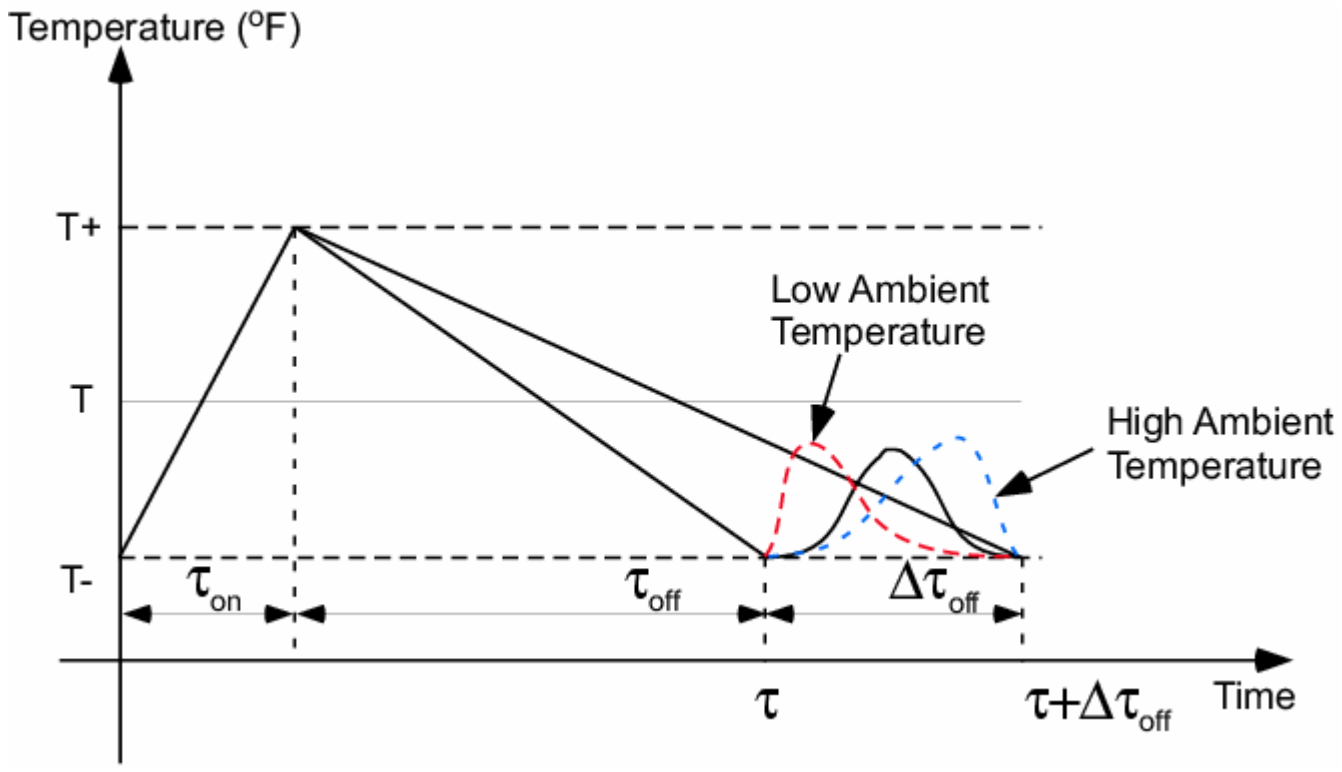

Fig. 7: Uncertainties in thermal characteristic curves

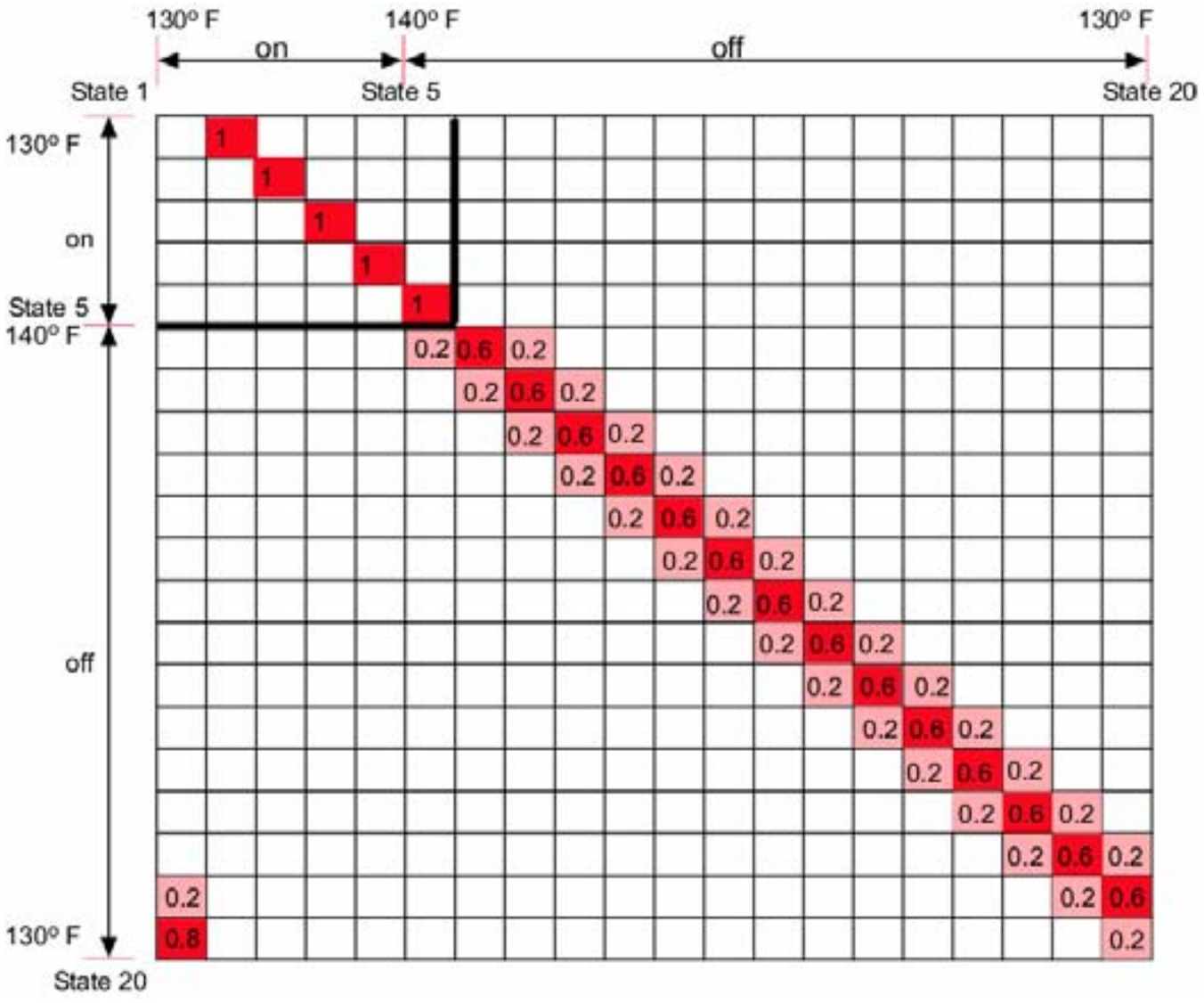

Fig. 8: Uncertainties in state transition matrix $P$

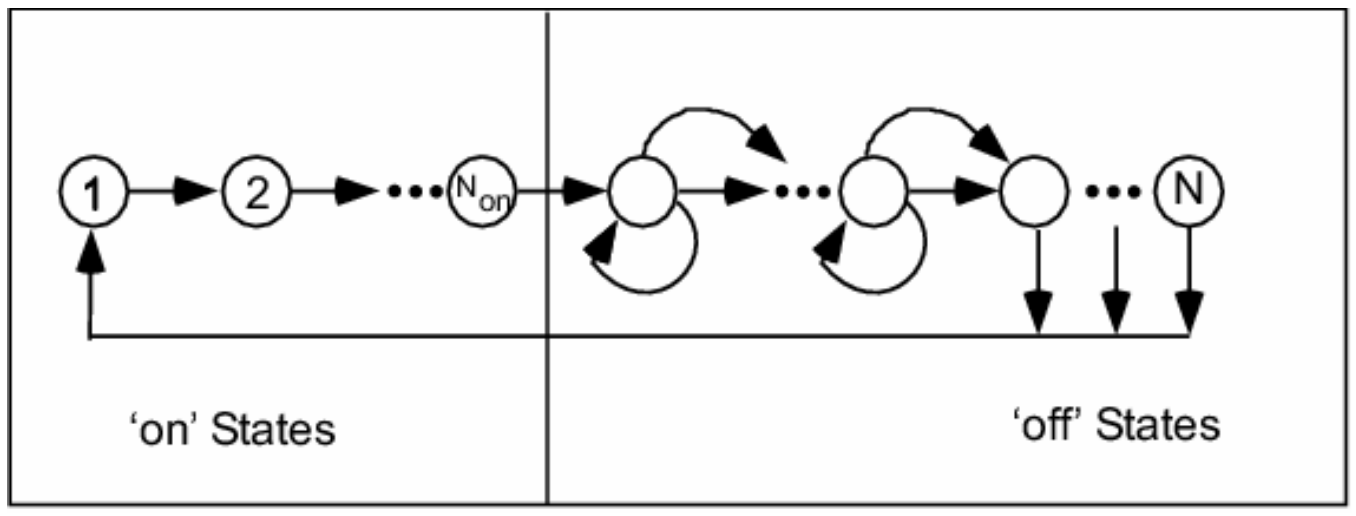

Fig. 9: A modified SQ model considering the uncertainties of the TCA cycling time

Let vector $X_k = (x_1, x_2, .., x_N)$ represent the number of units in each state at the $k^{th}$ time interval and $P$ represent the transition matrix [10] containing the transition probabilities $p_{i,j}$, which represents the probability of a unit moving from State $i$ to State $j$. We can then calculate the state evolution from the $k$-1 to the $k^{th}$ time interval using

$$X_k = X_{k-1}P \tag{4}$$

The aggregated power output $P_L$ is then calculated by

$$P_L = P_{ave}\sum_{m=1}^{N_{on}} x_m \tag{5}$$

where $P_{ave}$ is the average power output of a unit, $x_m$ is the number of units in an "on" State $m$, and $N_{on}$ is the number of the "on" states.

To account for the uncertainties, for units in State $i$, we have

$$\sum_{j=1}^{N} p_{i,j} = 1 \tag{6}$$

where $p_{i,j}$ follows certain PDF. For the 20-state transition probability matrix $P$ shown in Fig. 8, the unit residing in State 6 has probabilities 0.2, 0.6, and 0.2, to remain in State 6 or move forward to State 7 or 8. This shows that 20% of all units have slower or faster temperature decay rates than the majority, 60%.

In the next section, the impacts of customer random behaviors on aggregated TCA load curves are discussed.

## IV. Uncertainties Caused By Random Customer Behaviors

Customer behaviors can have a significant impact on the cycling time of TCAs. Based on load surveys, such as the End-Use Load and Consumer Assessment Program (ELCAP) data collected by the Bonneville Power Administration (BPA) [13], PDFs of diurnal, weekly, and annual load profiles can be obtained.

Fig. 10a shows a SQ model that takes customer hot water consumptions into account. Minor consumption is defined as a small amount of water draw, causing the water temperature to drop slightly. Because the temperature drops when the unit is on are not as significant as the temperature drops when the unit is off, the minor consumption can be modeled by adding $p_{i,i+m}$ ($m$ = 2, 3, …) entries for the "off" states and a $p_{i,i}$ entry for the "on" states in $P$ (Fig. 10b). A major consumption is defined as a large amount of hot water draw, which will cause the water temperature to drop below the lower temperature limit. The heater will then turn on. This kind of consumption can be modeled by putting a link between each state to State 1, as shown in Fig. 10b.

A transition matrix $P$ corresponding to the SQ model shown in Fig. 10b is

$$P=\begin{bmatrix} p_{1,1} & 1-p_{1,1} & & & & & & \\ p_{2,1} & p_{2,2} & 1-p_{2,1}-p_{2,2} & & & & & \\ \cdots & \cdots & \cdots & \cdots & \cdots & & & \\ p_{N_{on},1} & & p_{N_{on}N_{on}} & 1-p_{N_{on},1}-p_{N_{on}N_{on}} & & & & \\ \cdots & & & & \cdots & & \cdots & \\ p_{i,1} & & & & 1-p_{i,1}-p_{i,i+2}-\cdots & p_{i,i+2} & p_{i,i+3} & \cdots \\ 1 & & & & & & & \end{bmatrix}$$

where $p_{i,j}$ can be adjusted to account for the different consumption patterns. Note that $p_{i,j}$ is a function of time because the consumption of hot water varies by time-of-day.

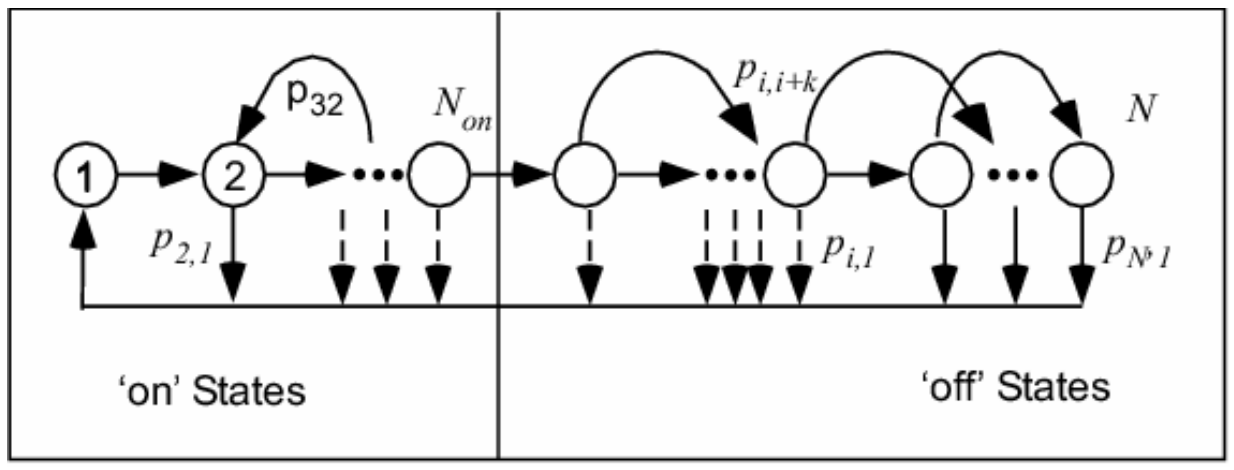

(a) The modified state queue

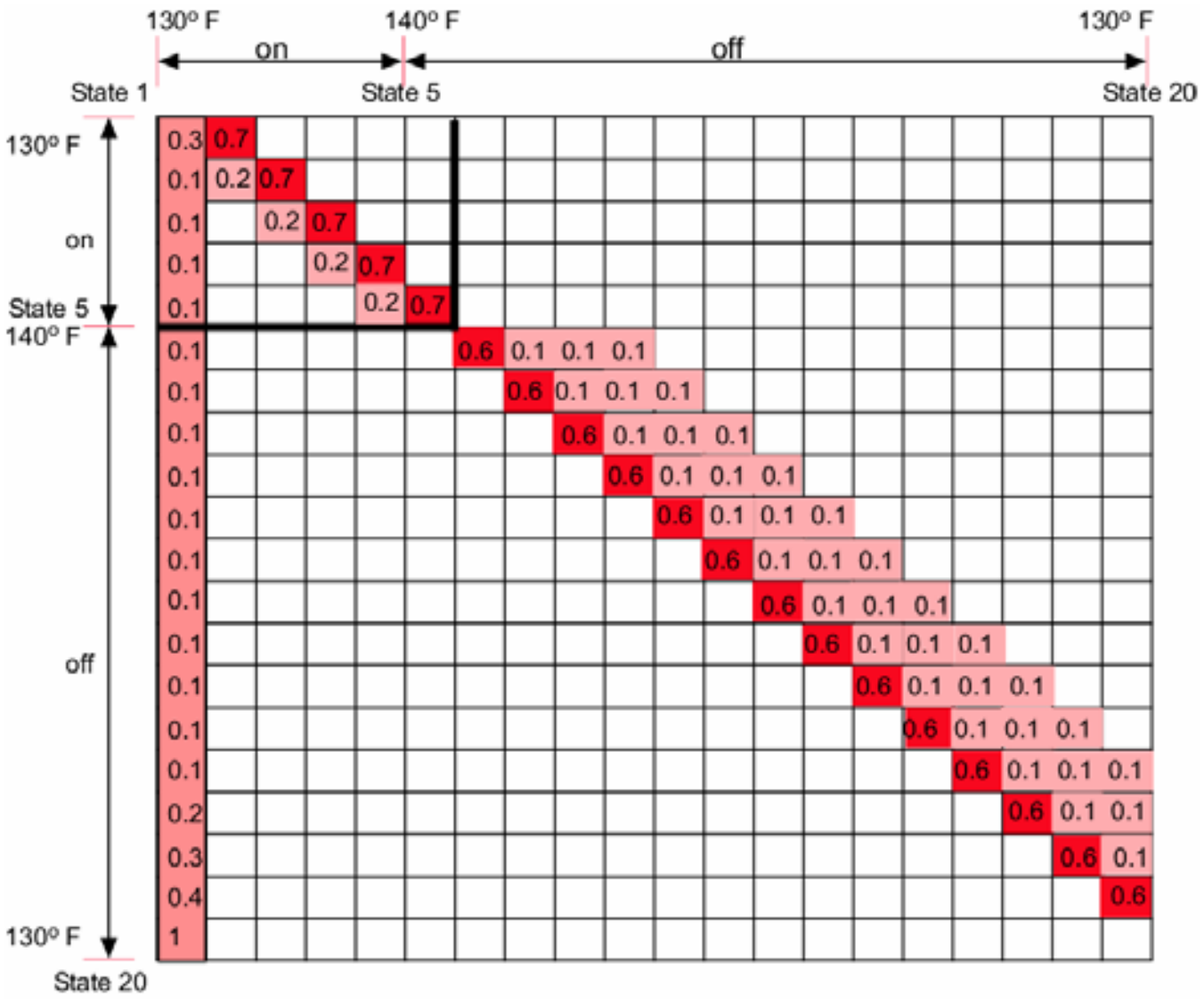

(b) The modified state transition matrix $P$

Fig. 10: The modified SQ model considering customer consumptions

To tune the state transition probability $p_{i,j}$, we use the data collected in ELCAP by BPA in our simulation. The customer consumptions are classified into two categories: major consumptions and minor consumptions. A major consumption includes behaviors such as washing dishes, taking showers, and washing clothes, which usually last for more than 5 minutes. A minor consumption includes behaviors such as washing hands and washing fruit, which usually last for 1 or 2 minutes.

Fig. 11 shows the ELCAP data curve versus the simulation data curve obtained using a 60-state SQ model with the probability curves shown in Fig. 12. The probability of major and minor consumptions during a winter weekday is calibrated by ELCAP data. To tune the $P$ matrix, we first estimate the probability of a minor consumption event based on consumption patterns. Then, we tune the major event probability to do the curve fitting. The calibration of the transition matrix $P$ to account for the parameter uncertainties follows an interval mathematics approach, because all variations of input parameters are simultaneously considered and reflected in our calibrated $P$ matrix with an assumption that all possible values for a parameter must lie somewhere within the assumed interval for the parameter [14]. Because an eigenvalue analysis on the $P$ matrix shows that the $p_{i,1}$ entries have a dominant impact on the steady-state distribution of the unit in each state for water heater case, the load profile follows the shape of the probability of major consumption events, as shown in Fig. 11 and Fig. 12.

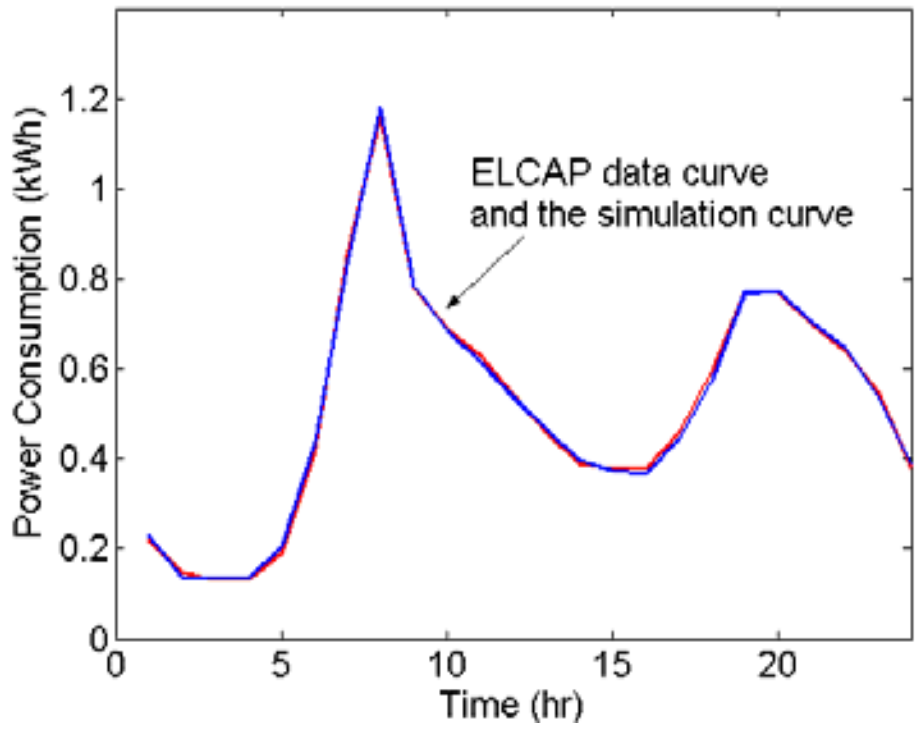

Fig. 11: Winter load profile for electric water heater load: ELCAP data versus simulation result

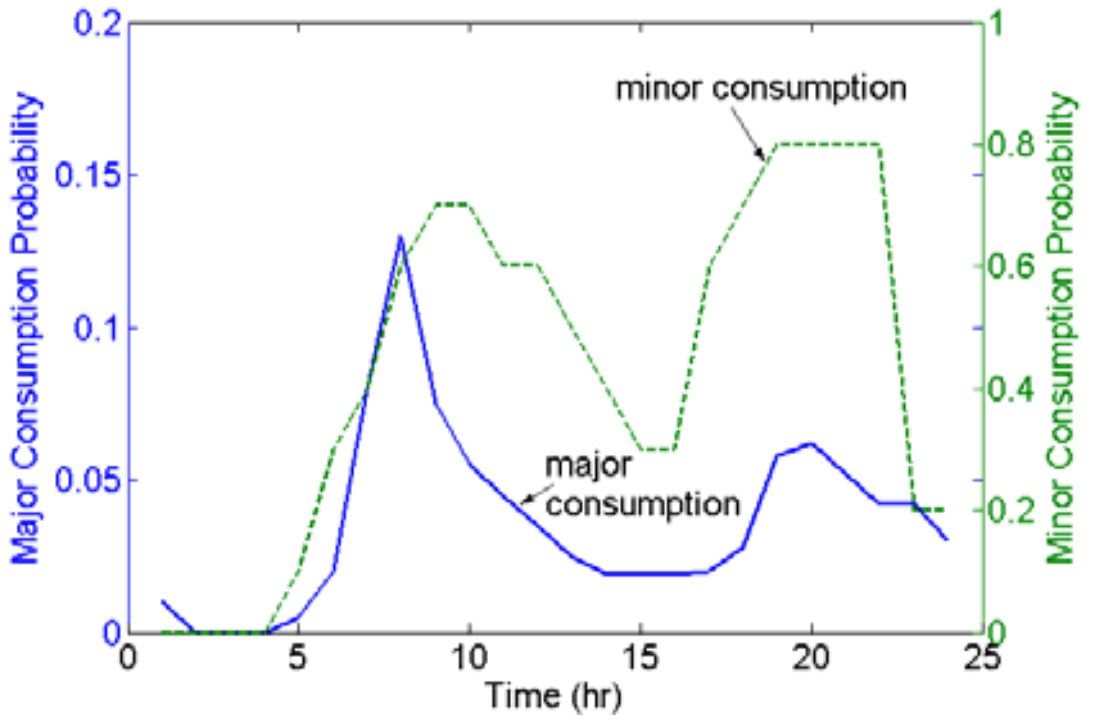

Fig. 12: The probabilities of major and minor hot water consumptions

## V. IMPACTS ON SETPOINT CHANGE RESPONSE OF TCAS

With a change in the setpoint, the upper temperature limit and the lower temperature limit of a TCA thermostat setting are shifted. Therefore, a unit may turn on/off accordingly, and electricity consumption is pre-consumed or deferred. However, the total energy consumed in 1 day remains the same.

### *A. The impact of uncertainties in load cycling times*

Assume a fixed dead band during a setpoint change. When the market price drops, we expect a setpoint to increase, as shown in Fig. 13, which results in moving all units to below the new lower temperature limit. Therefore, the units that are "off" before the change of setpoint will turn on. Using a 20-state SQ model with 5 "on" states as an example and considering 10,000 units distributed uniformly along the 20 states, the average number of "on" unit is 2,500. Considering only standby losses of the units, an oscillation on the aggregated load curve will be caused by a setpoint increase, as shown by the dotted line in Fig. 14. The detail of this oscillation has been analyzed in [8] using a TCA standby model.

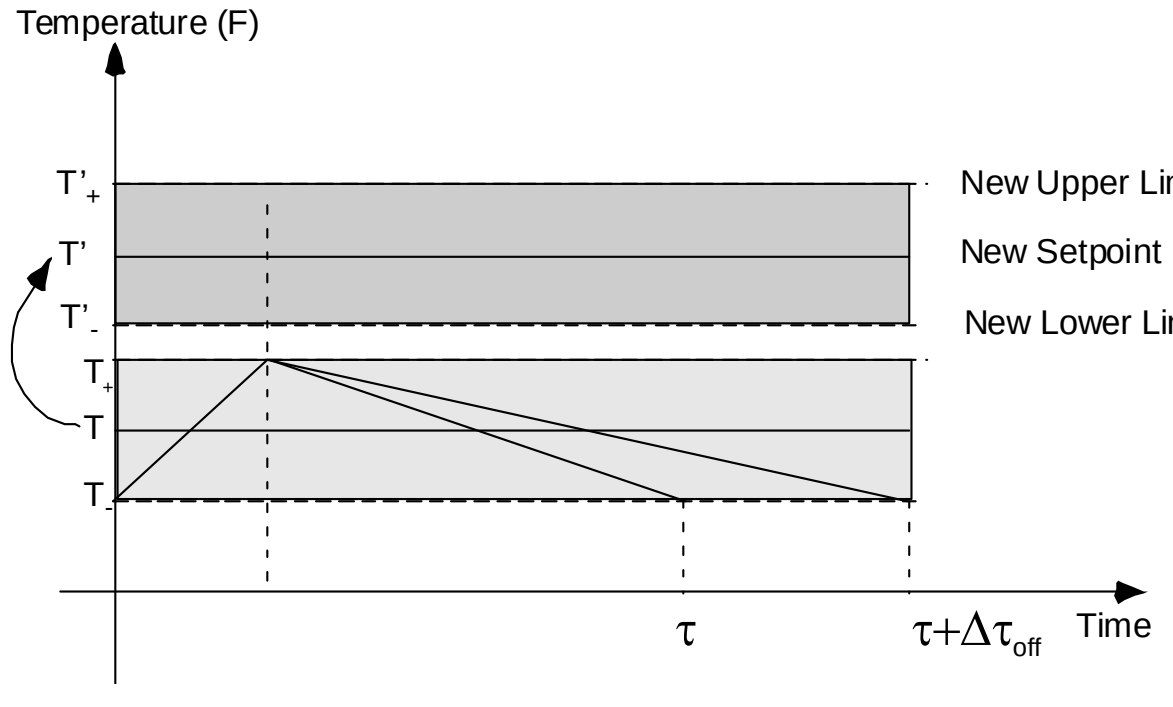

Fig. 13: A setpoint increase

In reality, the 10,000 units will not have exactly the same cycling time because of the differences in *UA*s and ambient temperatures. Using the modified transition matrix *P* shown in Fig. 8 to account for these uncertainties, the aggregated TCA response curve will then follow the solid line in Fig. 14. We note following observations:

- The initial peak is the same as that of the no-uncertainty case.
- The succeeding peaks are damped.
- The oscillation frequencies are different.

This is because the setpoint change basically will synchronize the units that are "off" with units that are "on" and form a dynamic oscillation with a frequency of $1/\tau$. The uncertainties in $\tau$ will cause a spectrum of oscillation frequency within $[1/\tau \; 1/(\tau+\tau_{off})]$. Therefore, the aggregated load peaks will be the highest initially. After that, because of the frequency differences, the aggregated load peaks will be less than the initial one. The more diverse the unit cycling times are, the faster the aggregated load peaks are damped.

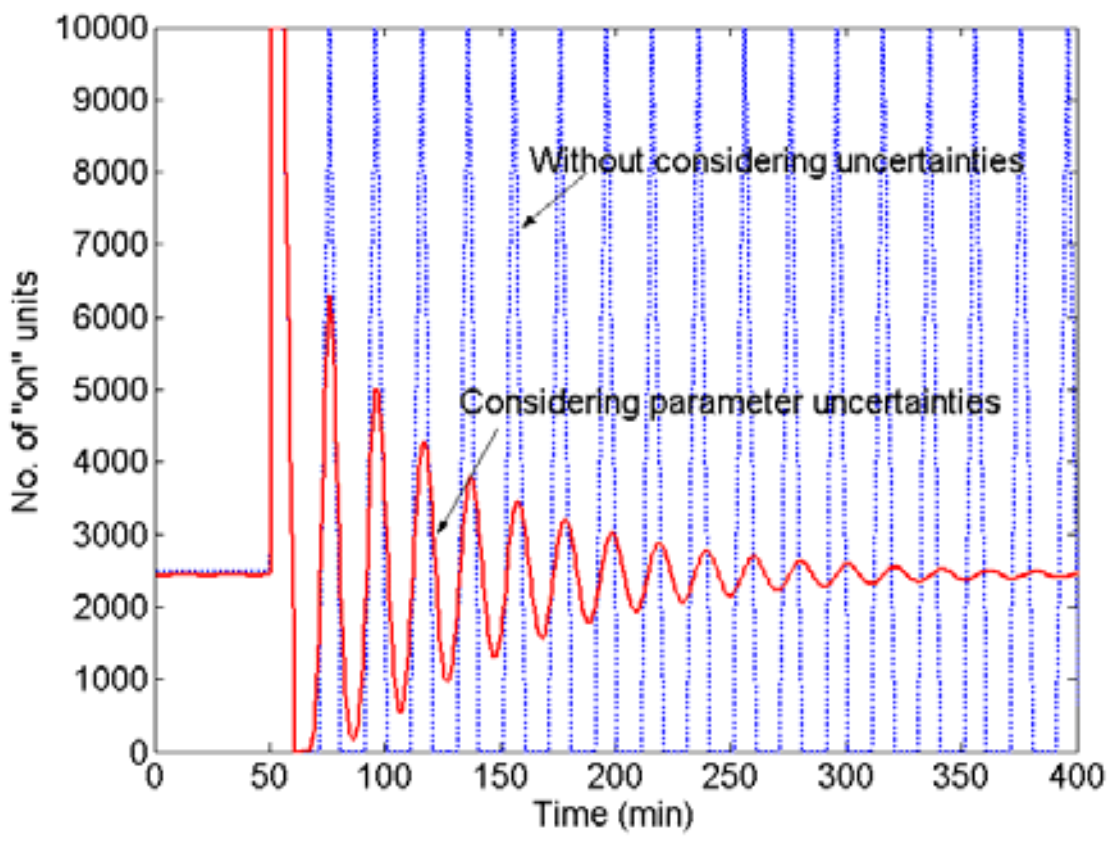

Fig. 14: The number of units in "on" state after a setpoint increase considering uncertainties in $\tau$

### *B. The impact of uncertainties in random load behaviors*

The uncertainties in load cycling times are uncertainties in aggregated load behaviors. For each individual TCA unit, the cycling time is deterministic for a given ambient temperature. However, load consumptions are different. Random demand may vary the cycling time of a TCA unit by randomly shortening the "off" cycle or by prolonging the "on" cycle, thereby increasing the effective duty cycle and bringing diversity to the water heater loads.

The 20-state water heater SQ model with 5 "on" states is again used for illustration purpose. The probability matrix shown in Fig. 10b is used to account for the customer consumptions. The setpoint increase response curve is shown in Fig. 15.

As discussed in Section IV, demand varies with respect to time. During night hours, both the major and the minor customer consumptions may be so infrequent that the probabilities of both can be considered to be zero. If a setpoint increase response is considered, we would expect the power consumption to follow curve 1 in Fig. 15. In the morning hours, people frequently use hot water and both the minor and major consumptions happen with high probabilities. If using the *P* matrix listed in Fig. 10b, the response will follow curve 2 in Fig. 15. During daytime, when people go to work, the hot water consumption may be less frequent. Assuming that the state transition probabilities for major consumptions change from 0.1 to 0.01 in Fig. 10, the response will follow curve 3 in Fig. 15. The setpoint decrease responses for the three scenarios are shown in Fig. 16.

There are several observations based on the results:

- Demand raises the average power output because the effective duty cycle is shorter than the standby model.
- Demand damps the load peaks after a setpoint increase and benefits the system by restoring the load diversity over time.
- As shown in Fig. 17, demand shortens the load shifting time obtained by lowering the setpoint. This

may have a negative impact on load shifting programs because the load may not be able to be shifted away from the peak-price periods totally.

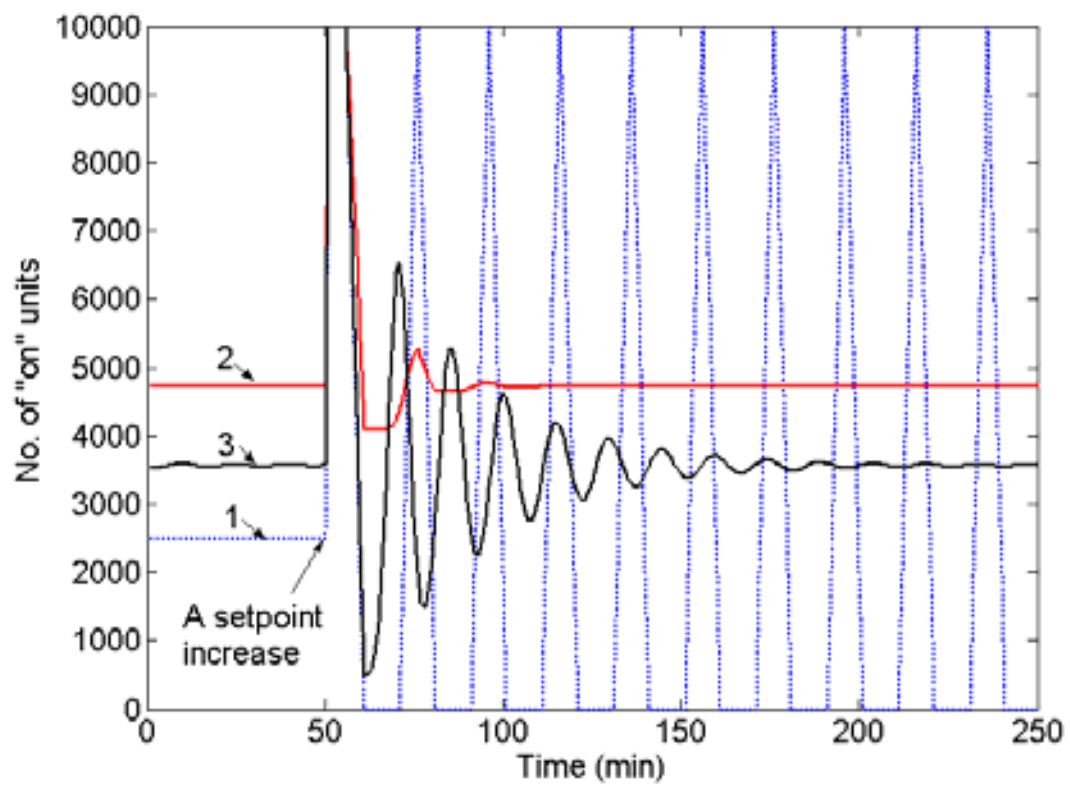

Fig. 15: The number of units in "on" state after a setpoint increase considering customer consumption

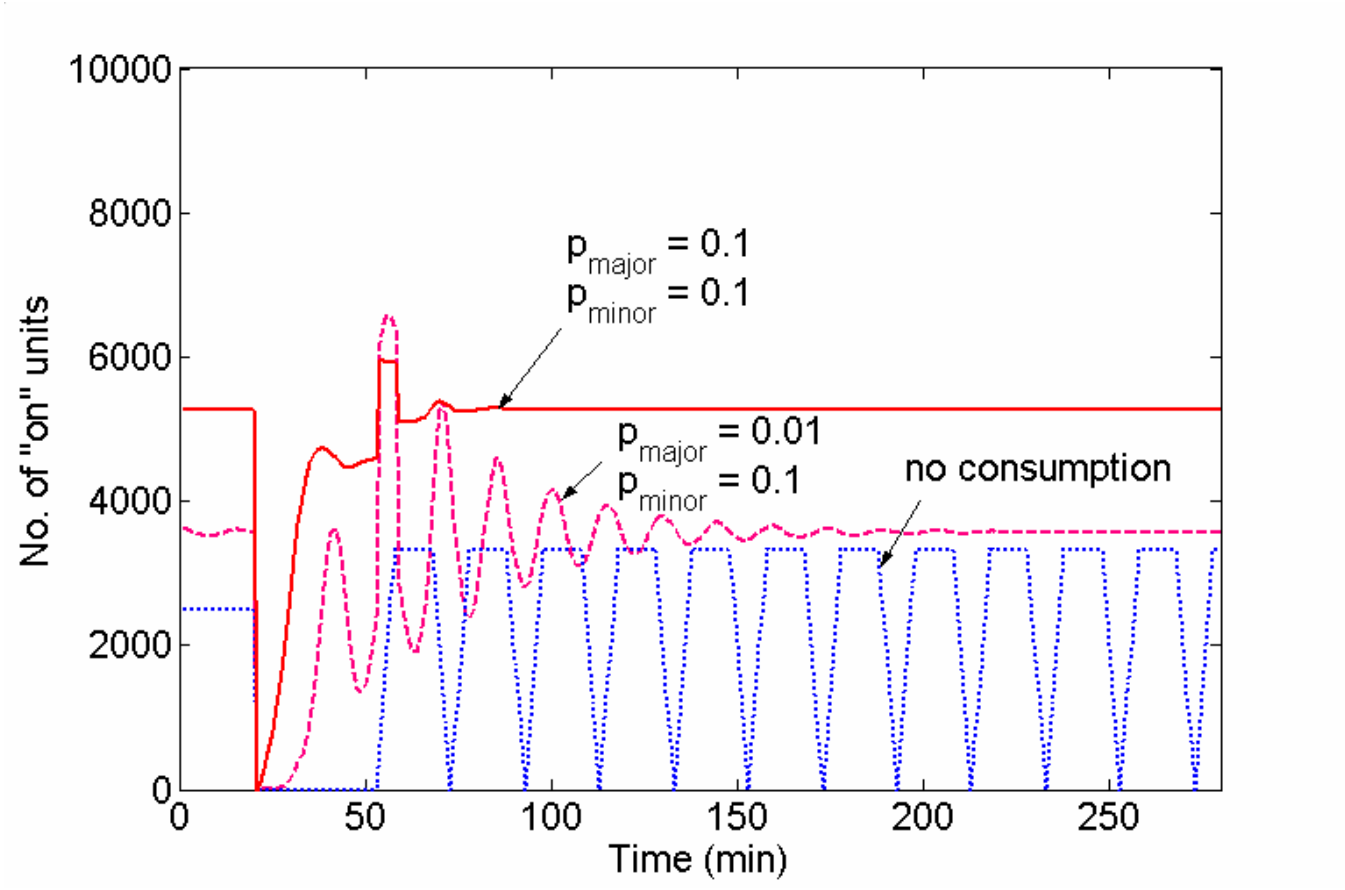

Fig. 16: The number of units in "on" state after a setpoint decrease considering the customer consumptions

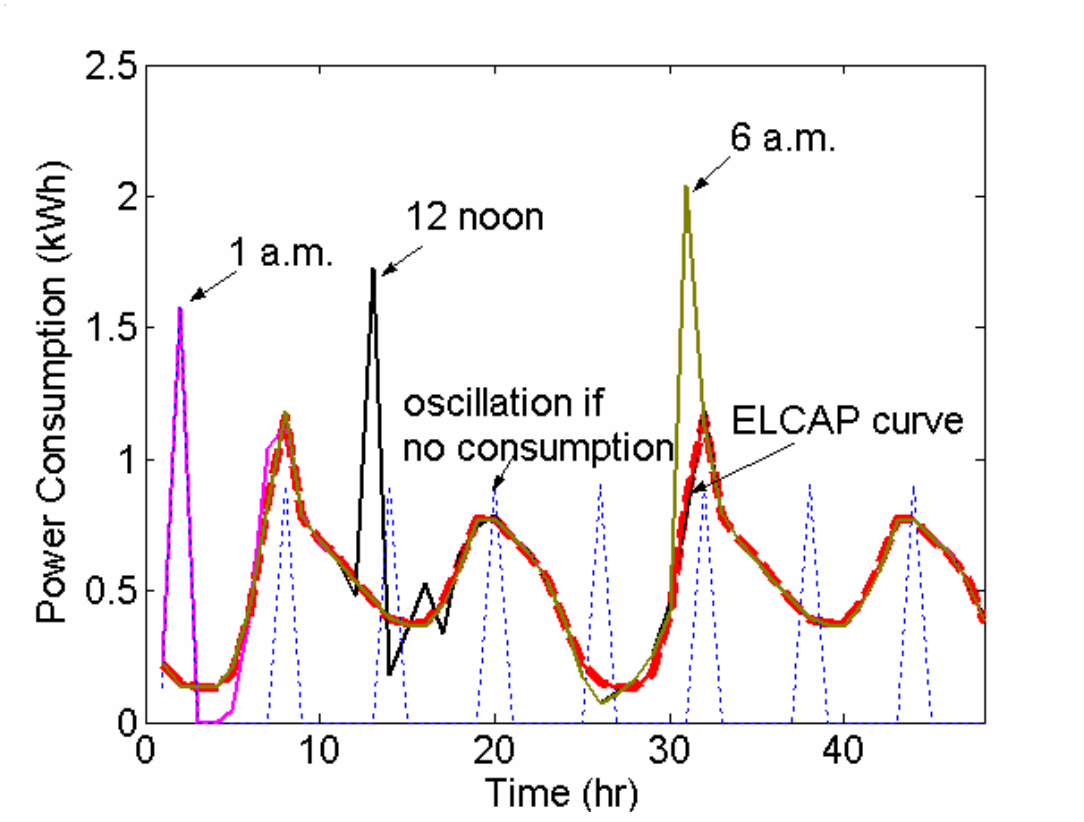

Fig. 17: Setpoint increase responses based on ELCAP data

Fig. 17 gives a simulation result using an ELCAP data tuned transition matrix. Assume that when prices drop, the setpoints of water heaters will be raised. Then, consider a price drop happens at 1 a.m. The dotted line shows the load oscillations after this price drop, if there is no customer consumption through a day. In real life, around 6 a.m., the increased hot water consumptions start to diffuse the synchronized loads and restore the original system diversity. If the price drop happens at 12 noon, when the minor consumptions are high but the major consumptions are low, then there is a small but observable oscillation in the load profile. This is because the many minor hot water consumptions damp the oscillation. If setpoints are raised at 6 a.m., when both the major and minor hot water consumptions are frequent, we do not observe this oscillation because the damping rate is high. By calculating the eigenvalues of the transition matrix $P$, one can then find out how fast load consumptions can damp the load oscillations caused by significant price drops or increases if the TCA setpoints are set to respond to price changes. The study can indicate whether or not distribution networks implementing such demand-side management strategies are overloaded under those circumstances and for how long.

### *C. The Combining Impacts*

Fig. 18 shows four cases of setpoint decrease responses. The solid lines are the responses considering both parameter uncertainties and the customer consumptions at different PDFs. The dotted lines show the cases considering only the customer consumptions.

The results indicate that, although the randomness of the load cycling times and the customer consumptions both contribute to restore load diversities after a setpoint response, customer consumptions have a dominant impact. This is because higher customer consumptions mean faster diffusion processes between states. A sensitivity analysis on the eigenvalues and eigenvectors of the transition matrix $P$ will also show that the higher the probability of customer consumptions, the higher the damping rates.

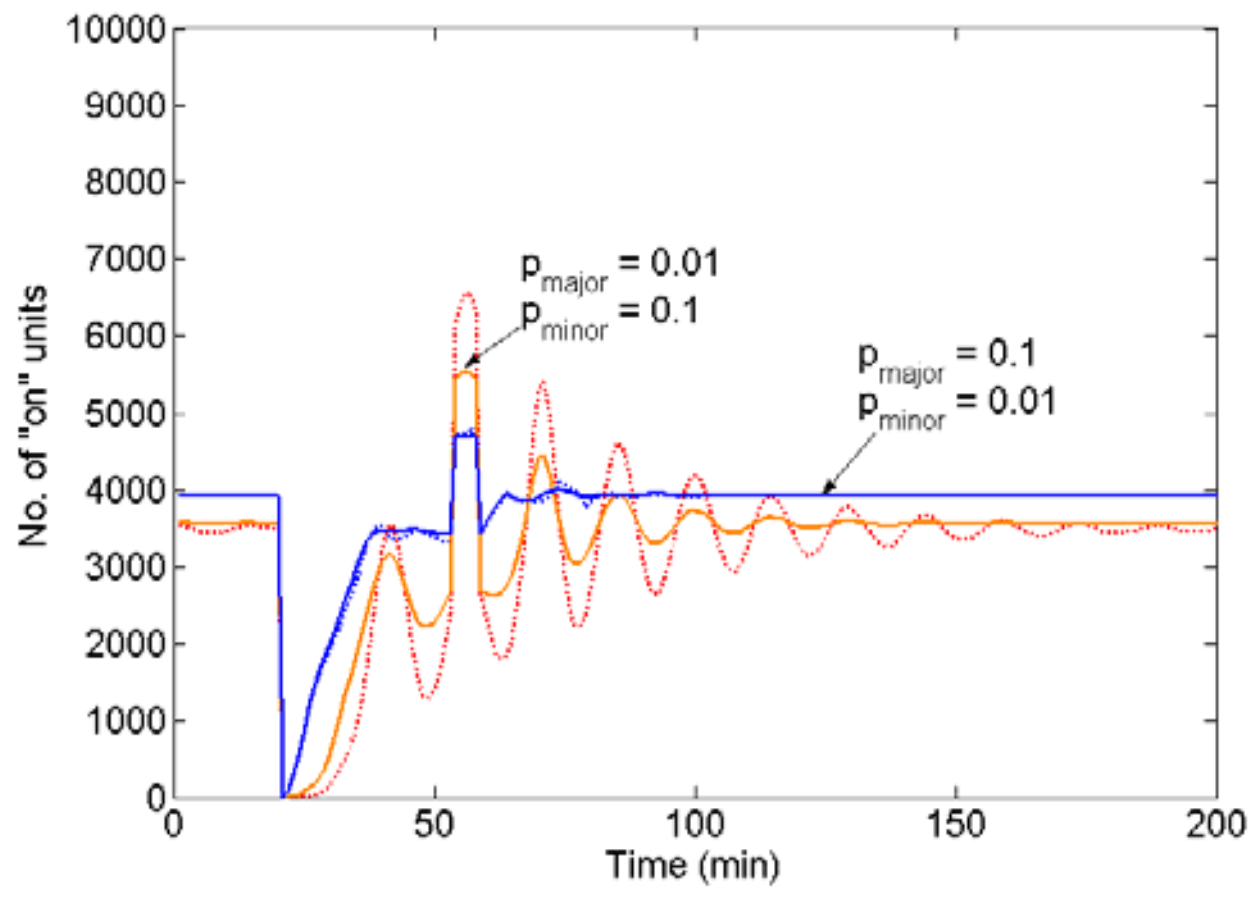

Fig. 18: The number of units in "on" state after a setpoint decrease (combining uncertainty impacts)

### *D. A comparison with physically unit-based simulation approach*

Power Distribution System Simulator (PDSS) [9] is a software simulation tool developed by Pacific Northwest National Laboratory (PNNL). PDSS is a detailed physically based simulation at the unit level, meaning that each water heater is simulated by a thermal model with its own set of parameters. Then, a load synthesis is done to get the aggregated load curve. Therefore, the uncertainties are accounted for at the unit level. The SQ model is a state-based model and it is an aggregated model. The uncertainties are accounted for at the system level. A comparison of results obtained by using software developed under the two different approaches is depicted in Fig. 19. The two software tools are first tuned by ELCAP data. Then they are used to simulate the load response to a setpoint decrease happening at 1 p.m. followed by a setpoint increase at 1 a.m. the next day.

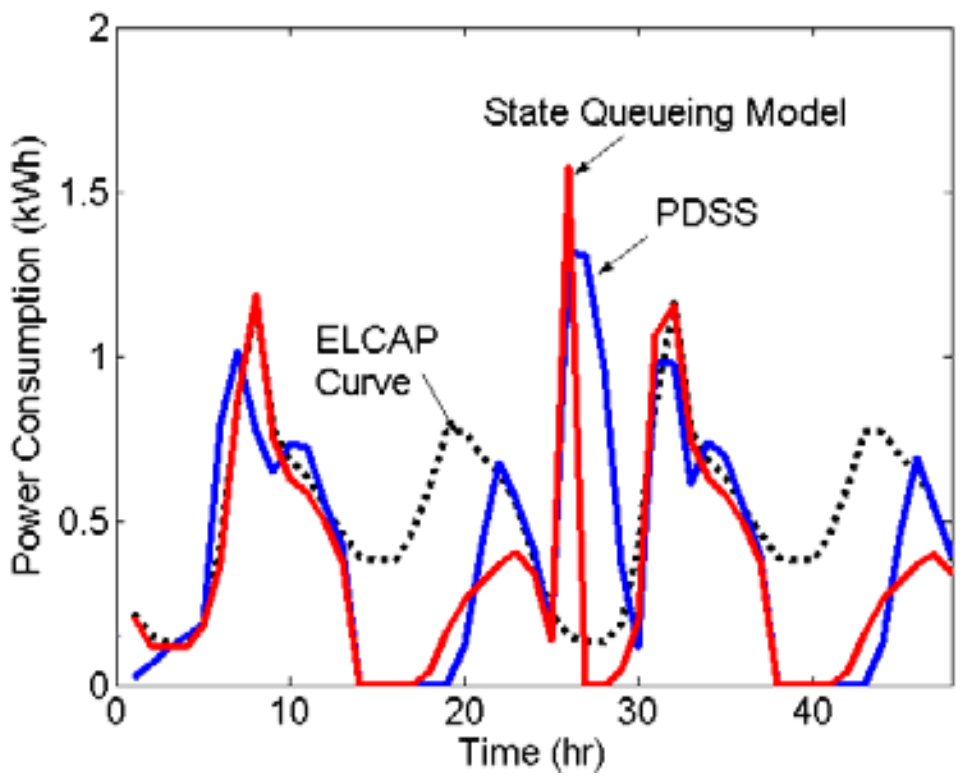

Fig. 19: A result comparison between PDSS and SQ model

From the result, we notice that

- The responses are similar, although the magnitude and the phase may have some deviations. Because PDSS accounts for uncertainties at the unit level, the aggregated load obtained is more diversified than the SQ model. The synchronized load peaks obtained by PDSS are lower.
- Compared with PDSS, the SQ model is easier to be tuned to follow a load curve exactly. This is because the input of the PDSS is at the unit level, and the aggregated load response at the system level is not fully controllable by the inputs.
- When the population size becomes larger, for example, if a feeder with thousands of water heaters is simulated, the SQ model computes much faster because the computation time of a SQ model is sensitive to the number of states, but not to the total number of units.

Because the two simulation approaches reach similar results with same inputs, it serves as a validation for both models. To summarize, the PDSS model shows more load diversity and has a better representation of peak load. The SQ model is computationally faster, although it yields a higher load peak with a faster damping rate.

## VI. CONCLUSION

This paper presents a method to include the modeling of uncertainties in the TCA load cycling times and random load consumptions by modifying the entries of the state transition matrix of a SQ model. From the load survey data, probability distribution functions of the cycling time and consumer behavior patterns can be obtained. State transition matrixes for each hour can then be tuned and used by the SQ model introduced in [8] to simulate the price response of loads aggregated by TCAs. The eigenvalues of the state transition matrix can be used to evaluate the system damping rate, which indicates the ability of the system to restore load diversity. Because the computation time of a SQ model is determined by the number of states used to model the TCA load cycling periods and the sparsity of the state transition matrix, it has computational advantages over unit-based simulation, thus holding promise as an useful equivalent load model for transmission level studies.

Our future work will focus on setting up the setpoint control functions, which govern setpoint responses to market prices. By evaluating the payments and resulting load profiles, the economic benefits and technical impacts of different control functions can then be evaluated.

## REFERENCES

[1] R. E. Bohn, *Spot Pricing of Public Utility Services*, Ph.D. dissertation, MIT, 1982. Also available as MIT Energy Laboratory Technical Report, MIT-EL 82-031, 1982.

[2] F. C. Schweppe, B. Daryanian, and R. D. Tabors, "Algorithms for a Spot Price Responding Residential Load Controller," *IEEE Transactions on Power Systems*, vol. 4, pp. 507–516, May 1989.

[3] B. Daryanian, R. E. Bohn, and R. D. Tabors, "Optimal Demand-side Response to Electricity Spot Prices for Storage-type Customers," *IEEE Trans. on Power Systems*, vol. 4, pp. 897–903, Aug. 1989.

[4] C. W. Gellings and R. W. Taylor, "Electric Load Curve Synthesis – A Computer Simulation of an Electric Utility Load Shape," *IEEE Trans. on Power Apparatus and Systems*, vol. PAS-100, No. 1, Jan. 1981.

[5] M. L. Chan, E. N. Marsh, J. Y. Yoon, G. B. Ackerman, and N. Stoughton, "Simulation-based Load Synthesis Methodology for Evaluating Load-management Programs," *IEEE Trans. on Power Apparatus and Systems*, vol. PAS-100, pp. 1771-1778, April 1981.

[6] A. Pahwa and C. W. Brice III, "Modeling and System Identification of Residential Air Conditioning Load," *IEEE Trans. on Power Apparatus and Systems*, vol. PAS-104, No. 6, June 1985.

[7] S. Ihara and F. C. Schweppe, "Physically Based Modeling of Cold Load Pickup," *IEEE Trans. on Power Apparatus and Systems*, vol. PAS-100, No. 9, Sept. 1981.

[8] N. Lu and D. P. Chassin, "A State Queueing Model of Thermostatically Controlled appliances," *IEEE trans. on power systems*, vol. 19, No.3, pp. 1666-1673, Aug. 2004.

[9] R. T. Guttromson, D. P. Chassin, and S. E. Widergren, "Residential Energy Resource Models for Distribution Feeder Simulation," *Proc. of 2003 IEEE PES General Meeting*, Toronto, Canada, pp. 108-113, 2003.

[10] D. Gross and C. M. Harris, *Fundamentals of Queueing Theroy*, 3rd ed., John Wiley & Sons. Inc., 1998.

[11] R. G. Pratt and Z. T. Taylor, "Development and Testing of an Equivalent Thermal Parameter Model of Commercial Buildings from Time-Series End-Use Data," Pacific Northwest Laboratory, Richland, WA, Apr. 1994.

[12] Z. T. Taylor and R. G. Pratt, "The Effects of Model Simplifications on Equivalent Thermal Parameters Calculated from Hourly Building Performance Data," in *Proceedings of the 1988 ACEEE Summer Study on Energy Efficiency in Buildings*, pp. 10.268-10.285, 1988.

[13] F. J. Peterson, J. E. Patton, M. E. Miller, R. A. Gillman, W. M. Warwick, and W. F. Sandusky, "End-Use Load and Consumer Assessment Program," *Energy and Buildings*, ISSN 0378-7788, vol. 19, No. 3, 1993.

[14] R.P. Broadwater, H. E. Shaalan, W. J. Fabrycky, R. E. Lee, "Decision evaluation with interval mathematics: a power distribution system case study," *IEEE Transactions on power delivery*, vol. 9, Issue: 1, pp. 59 – 67, Jan. 1994.

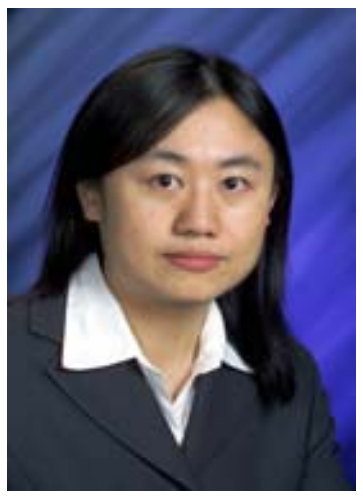

**Ning Lu** (M'98) received her B.S.E.E. from Harbin Institute of Technology, Harbin, China, in 1993, and her M.S. and Ph.D. degrees in electric power engineering from Rensselaer Polytechnic Institute, Troy, New York, in 1999 and 2002, respectively. Her research interests are in the modeling and analysis of deregulated electricity markets. Currently, she is a research engineer with the Energy Science & Technology Division, Pacific Northwest National Laboratory, Richland, WA. She was with Shenyang Electric Power Survey and Design Institute from 1993 to 1998.

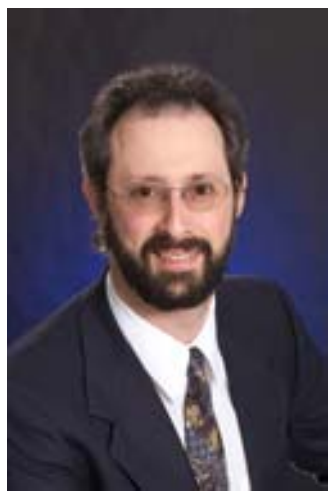

**David P. Chassin** received his B.S. of Building Science from Rensselaer Polytechnic Institiute in Troy, New York. He is staff scientist with the Energy Science & Technology Division at Pacific Northwest National Laboratory, where he has worked since 1992. He was Vice-President of Development for Image Systems Technology from 1987 to 1992, where he pioneered a hybrid raster/vector computer aided design (CAD) technology called CAD Overlay™. He has experience in the development of building energy simulation and diagnostic systems, leading the development of Softdesk Energy and DOE's Whole Building Diagnostician. He has served on the International Alliance for Interoperability's Technical Advisory Group, and chaired the Codes and Standards Group. His recent research focuses on emerging theories of complexity as they relate to high-performance simulation and modeling.

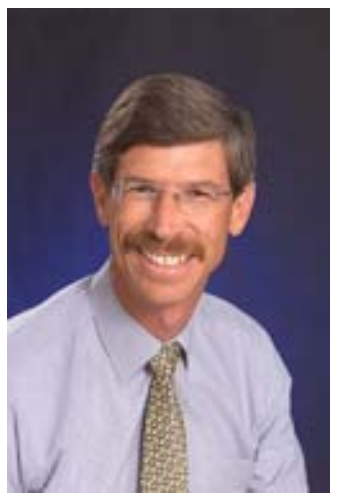

**Steve Widergren** (M'1978, SM'1992) received his B.S.E.E. (1975) and M.S.E.E. (1978) from the University of California, Berkeley. He works at the Pacific Northwest National Laboratory, where he contributes to the research and development of new solutions for the economic and reliable operation of power systems. Prior to joining PNNL, he was with ALSTOM ESCA, a bulk transmission energy management system supplier, where he designed software solutions for system operations and championed the establishment of an integrated suite of energy management system software products. He has also held power engineering positions at American Electric Power and Pacific Gas & Electric (as an intern). He is vice-chair of the PES Energy Control Center Subcommittee and a member of the IEEE SCC 21 1547.3 standards group on monitoring, information exchange, and control of distributed resources.